\documentclass[aps, prl, superscriptaddress, citeautoscript, amsmath, amssymb, showkeys, reprint]{revtex4-2}

\usepackage{xr}
\makeatletter
\newcommand*{\addFileDependency}[1]{
  \typeout{(#1)}
  \@addtofilelist{#1}
  \IfFileExists{#1}{}{\typeout{No file #1.}}
}
\makeatother

\newcommand*{\myexternaldocument}[1]{
    \externaldocument[S]{#1}
    \addFileDependency{#1.tex}
    \addFileDependency{#1.aux}
}

\myexternaldocument{Supplement} 

\usepackage{graphicx}
\usepackage{dcolumn}
\usepackage{siunitx}
\usepackage{bm}
\usepackage{hyperref}
\usepackage{xargs}                      
\usepackage[pdftex,dvipsnames]{xcolor}  
\usepackage[colorinlistoftodos,prependcaption,textsize=tiny]{todonotes}
\newcommandx{\unsure}[2][1=]{\todo[linecolor=red,backgroundcolor=red!25,bordercolor=red,#1]{#2}}
\newcommandx{\change}[2][1=]{\todo[linecolor=blue,backgroundcolor=blue!25,bordercolor=blue,#1]{#2}}
\newcommandx{\info}[2][1=]{\todo[linecolor=OliveGreen,backgroundcolor=OliveGreen!25,bordercolor=OliveGreen,#1]{#2}}
\newcommandx{\improvement}[2][1=]{\todo[linecolor=Plum,backgroundcolor=Plum!25,bordercolor=Plum,#1]{#2}}
\newcommandx{\thiswillnotshow}[2][1=]{\todo[disable,#1]{#2}}

\begin{document}

\title{Approaching the intrinsic properties of moir\'e structures using AFM ironing}


\author{Swaroop Kumar Palai}
\affiliation{Laboratoire National des Champs Magn\'etiques Intenses, UPR 3228, CNRS-UGA-UPS-INSA, Grenoble and Toulouse, France}

\author{Mateusz Dyksik}
\affiliation{Department of Experimental Physics, Faculty of Fundamental Problems of Technology, Wroc{\l}aw University of Science and Technology, Wroc{\l}aw, Poland}

\author{Nikodem Sokolowski}
\affiliation{Laboratoire National des Champs Magn\'etiques Intenses, UPR 3228, CNRS-UGA-UPS-INSA, Grenoble and Toulouse, France}

\author{Mariusz Ciorga}
\affiliation{Department of Experimental Physics, Faculty of Fundamental Problems of Technology, Wroc{\l}aw University of Science and Technology, Wroc{\l}aw, Poland}

\author{Estrella Sánchez Viso}
\affiliation{Materials Science Factory, Instituto de Ciencia de Materiales de Madrid (ICMM-CSIC), Madrid, E-28049, Spain}

\author{Yong Xie}
\affiliation{Materials Science Factory, Instituto de Ciencia de Materiales de Madrid (ICMM-CSIC), Madrid, E-28049, Spain}

\author{Alina Schubert}
\affiliation{Materials Science Factory, Instituto de Ciencia de Materiales de Madrid (ICMM-CSIC), Madrid, E-28049, Spain}

\author{Takashi Taniguchi }\affiliation{International Center for Materials Nanoarchitectonics,  National Institute for Materials Science, Tsukuba, Ibaraki 305-004, Japan}

\author{Kenji Watanabe  }\affiliation{Research Center for Functional Materials, National Institute for Materials Science, Tsukuba, Ibaraki 305-004, Japan}

\author{Duncan K.\ Maude}
\affiliation{Laboratoire National des Champs Magn\'etiques Intenses, UPR 3228, CNRS-UGA-UPS-INSA, Grenoble and Toulouse, France}

\author{Alessandro Surrente}
\affiliation{Department of Experimental Physics, Faculty of Fundamental Problems of Technology, Wroc{\l}aw University of Science and Technology, Wroc{\l}aw, Poland}

\author{Micha{\l} Baranowski}
\affiliation{Department of Experimental Physics, Faculty of Fundamental Problems of Technology, Wroc{\l}aw University of Science and Technology, Wroc{\l}aw, Poland}

\author{Andres Castellanos-Gomez}
\affiliation{Materials Science Factory, Instituto de Ciencia de Materiales de Madrid (ICMM-CSIC), Madrid, E-28049, Spain}

\author{Carmen Munuera}\email{cmunuera@icmm.csic.es}
\affiliation{Materials Science Factory, Instituto de Ciencia de Materiales de Madrid (ICMM-CSIC), Madrid, E-28049, Spain}

\author{Paulina Plochocka}\email{paulina.plochocka@lncmi.cnrs.fr}
\affiliation{Laboratoire National des Champs Magn\'etiques Intenses, UPR 3228, CNRS-UGA-UPS-INSA, Grenoble and Toulouse, France}
\affiliation{Department of Experimental Physics, Faculty of Fundamental Problems of Technology, Wroc{\l}aw University of Science and Technology, Wroc{\l}aw, Poland}

\date{\today}
        
\begin{abstract}
Stacking monolayers of transition metal dichalcogenides (TMDs) 
has led to the discovery of a plethora of new exotic phenomena, 
resulting from moir\'e pattern formation. Due to the atomic thickness and high surface-to-volume ratio of heterostructures, 
the interfaces play a crucial role. Fluctuations in the interlayer distance affect interlayer coupling and  moir\'e effects. 
Therefore, to access the intrinsic properties of the TMD stack, it is essential to obtain a clean and uniform interface between the layers. Here, we show that this is {achieved} by ironing with the tip of an atomic force microscope. This post-stacking procedure dramatically improves the homogeneity of the interfaces, which is reflected {in the optical response of the interlayer exciton}. We demonstrate that ironing improves the layer coupling, enhancing moir\'e effects and reducing disorder. This is crucial for the investigation of TMD heterostructure physics, which currently suffers from low reproducibility.
\end{abstract}


                             
\maketitle

Van der Waals{(vdW)} crystals are considered key enablers of the next generation electronic and optoelectronic devices with improved performance 
\cite{radisavljevic2011single, duan2015two, wei2018various, ahn20202d, schaibley2016valleytronics, buscema2014fast}. 
The absence of lattice matching constraints 
enables new paradigms of material engineering, in which semiconductors, metal, superconductors, insulators or topological insulators can be seamlessly combined \cite{geim2013van}.

\begin{figure*}[ht]
    \centering
    \includegraphics{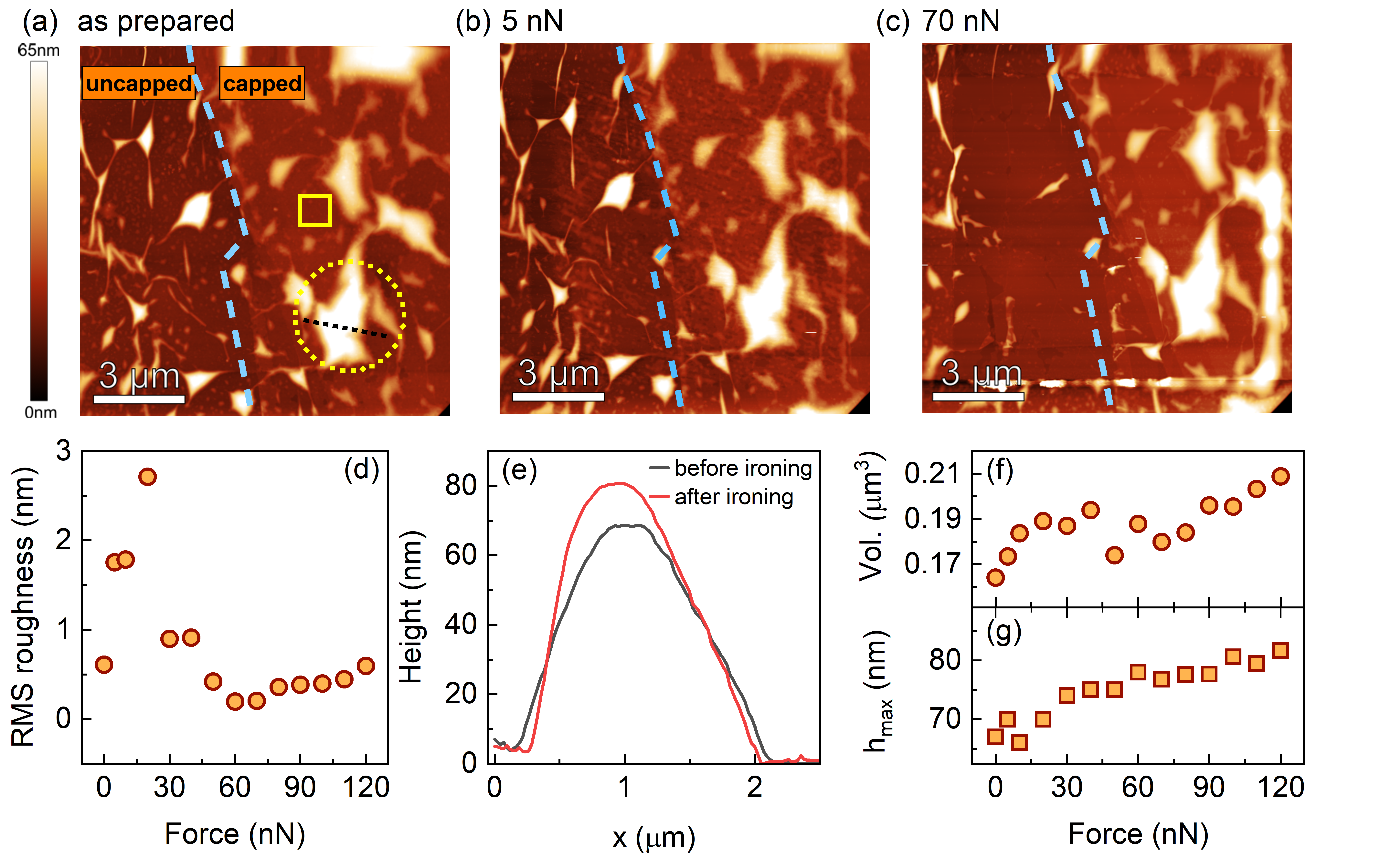}
    \caption{{Topographic atomic force microscopy (AFM) images in the dynamic mode of the (a) as-prepared samples and after ironing with (b) 5 and (c) \SI{70}{\nano\newton}, demonstrating the impact of incremental ironing forces on the sample. (d) Root mean square (RMS) values as a function of the ironing forces exerted on the region of interest (\SI{1}{\micro\metre^2}) indicated by the square in (a). (e) Height profile of the bubble of interest (dotted circle in (a)) taken along the black dash line cross-section, scanned in dynamic mode,  before and after ironing with \SI{120}{\nano\newton}. (f) Volume and (g) height of the bubble of interest as a function of the applied force during ironing.}}
    \label{fig:Fig1_afm_monolayer}
\end{figure*}

\begin{figure*}[ht]
    \centering
    \includegraphics[width=1\linewidth]{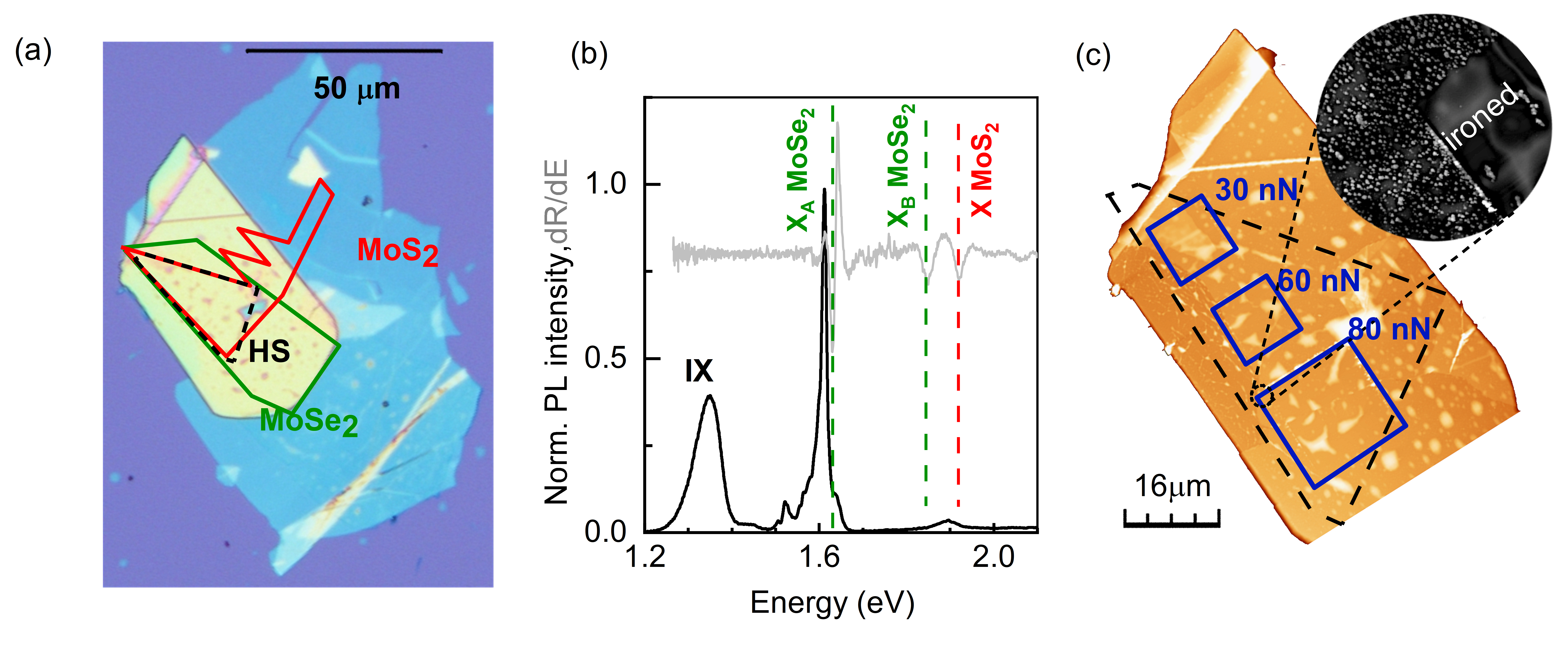}
    \caption{{(a) Optical micrograph of sample A, highlighting MoS$_2$ and MoSe$_2$ flakes (red and green outlines, respectively) and the heterostructure region (black outline). (b) Photoluminescence and reflectance spectra obtained at 4.5K, featuring inter- and intralayer exciton resonances (indicated). (c) Topographic AFM scan in dynamic mode revealing regions ironed with varying forces, with a close-up view of the ironed area (\SI{80}{\nano\newton}) in the inset. }}
    \label{fig:Fig2_optical}
\end{figure*}

{Fabricating vdW stacks has revolutionized the investigation of graphene \cite{dean2010boron, yankowitz2012emergence, chen2019signatures} and transition metal dichalcogenides (TMD) \cite{cadiz2017excitonic}. Encapsulating these stacks with hexagonal boron nitride (hBN) \cite{dean2010boron,cadiz2017excitonic, tongay2013broad, raja2019dielectric}} {has} provided access to their intrinsic properties, which are otherwise hidden by 
charged impurities and dielectric disorder \cite{ raja2019dielectric, cadiz2017excitonic}. Simultaneously, the precise control 
relative twist angle in bilayer graphene or TMDs led to the discovery 
of new phenomena such as superconductivity \cite{cao2018unconventional}, 
or long-lived interlayer optical transitions with controllable selection rules \cite{tran2019evidence, seyler2019signatures}.

In this perspective, the quality of the interfaces between the layers in the stack is of utmost importance to unveil their intrinsic properties, 
which are highly sensitive to their surroundings \cite{raja2017coulomb}
. Moreover, the mechanical exfoliation of bulk crystals \cite{castellanos2014isolation} unavoidably entails some form of interface disorder, which include charged impurities, traps \cite{yu2017analyzing}, wrinkles or bubbles \cite{gasparutti2020clean, rosenberger2018nano, han2021criteria, chen2021tip}. {These localized strain gradients impact profoundly the electronic properties of layered materials \cite{harats2020dynamics,wang2021visualizing,levy2010strain}, and are considered to be one of the main sources of disorder in high-quality samples \cite{couto2014random}.} Providing long-range homogeneity of the interface represents an ongoing challenge, and the lack thereof explains the considerable variations of the optoelectronic properties of nominally identical heterostructures \cite{tran2019evidence, seyler2019signatures, hanbicki2018double}. {Reports on the interlayer exciton in TMD heterobilayers show significant variation in the number of peaks and their energy, with a 
 range of up to 110 meV \cite{hanbicki2018double,rivera2015observation,ciarrocchi2019polarization,forg2021moire}. Even in high-quality monolayers, the energy of the exciton transition can vary as much as 10 meV  \cite{cadiz2017excitonic}.}


{Various preparation techniques such as transfer in inert atmosphere \cite{jung2019transferred} and annealing \cite{kretinin2014electronic} have been effective in minimizing defects in layered materials. Lately, AFM ironing \cite{rosenberger2018nano,chen2021tip, kim2019reliable, goossens2012mechanical} has proven to further reduce these defects and enhance optical properties by enabling the observation of interlayer exciton at room temperature \cite{rosenberger2018nano}, bilayer Raman peaks \cite{hanbicki2018double}, and moir{\'e} pattern reconstructions \cite{rosenberger2020twist}. However, systematic studies of the impact of AFM ironing on the optical properties of the interlayer excitons are needed.}



Here, we demonstrate that AFM ironing 
improves the interface quality and, as a consequence, 
the optical properties of hBN-encapsulated MoS$_2$/MoSe$_2$ heterostructure \cite{baranowski2017probing, zhang2018moire,lin2023remarkably}. 
{
We probe the quality of the interface by optical methods, focusing on the interlayer exciton transitions \cite{rivera2015observation,rivera2016valley,nagler2017interlayer,miller2017long,tran2019evidence, seyler2019signatures}, which are highly sensitive to the inter-layer distance. We demonstrate that ironing improves the layer coupling, enhancing moir\'e effects and reducing disorder. }
{We begin our investigation by ironing a MoS$_2$ monolayer deposited on an hBN substrate and partially capped with another hBN flake (see Tab.\ \ref{Sfig:table} in the Supplementary Information (SI)). With an AFM, we perform an ironing procedure by scanning selected areas with the tip in contact mode using different pressing forces. We study the evolution of the surface morphology as a function of the applied force. We then} investigate {using optical spectroscopy} three {hBN-encapsulated MoS$_2$/MoSe$_2$ heterobilayers}: Samples A and B with a nominal twist angle of \SI{60}{\degree} and sample C with \SI{0}{\degree} alignment, which facilitates strong interlayer exciton emission \cite{seyler2019signatures, nayak2017probing} (see table \ref{Sfig:table} in the SI for sample details). Within the ironed areas, the surface homogeneity can be greatly enhanced, with a reduction in the root mean square (RMS) roughness compared to the non-ironed regions. The increased interface quality of the heterostructure gives rise to a more uniform PL spectrum from the ironed parts, whose properties correlate with the applied ironing force. The high reproducibility of the properties of van der Waals stacks we demonstrate here qualifies AFM ironing as a strategy for the post-fabrication treatment of van der Waals heterostructures, providing access to their intrinsic properties.  

{We investigate the effects of the AFM ironing by examining a MoS$_2$ monolayer exfoliated on an hBN substrate and only partially capped by another hBN flake. In Fig.\ \ref{fig:Fig1_afm_monolayer}(a), we show the AFM image of the as-fabricated sample measured in dynamic mode, prior to any AFM treatment. The area we selected straddles the edge of the top hBN flake, highlighted by the blue dashed line visible in Fig.\ \ref{fig:Fig1_afm_monolayer}(a-c). A large number of bubbles and wrinkles can be noted, resulting from the stamping process of the exfoliated flakes \cite{gasparutti2020clean,rosenberger2018nano,han2021criteria,chen2021tip}. We then scan an area of \SI{10}{\micro\metre} by \SI{10}{\micro\metre}  in contact mode, by applying subsequently increasing forces ranging from \SI{5}{\nano\newton} to \SI{120}{\nano\newton}. After each scan, the initial area (\SI{12}{\micro\metre} by \SI{12}{\micro\metre}) was imaged in dynamic mode, to verify the effects of the scans in contact mode. Based on the effects on both the surface morphology and the concentration of bubbles, we identify two ranges of forces. In the low force limit (applied force smaller than \SI{20}{\nano\newton}), the debris and adsorbates present on the surfaces are initially moved and clustered, forming ripple-like features on the surface of the sample in the direction parallel to the fast scan, before being completely swept away from the scanned area by the AFM tip. An example of adsorbate clusters is visible in Fig.\ \ref{fig:Fig1_afm_monolayer}(b), where oblique bright lines can be seen in the scanned area, and is highlighted in Fig.\ \ref{Sfig:AFM_Debris_cluster} of the SI. This, in turn, induced an increase of the roughness in the low force limit, which we report in Fig.\ \ref{fig:Fig1_afm_monolayer}(d). Despite the observed changes, it is important to note that the larger defects, including bubbles and wrinkles, remain unaffected, and no other discernible alterations to the surface morphology are apparent. Their number, shape and size do not change as long as the applied force is small enough. However, when the applied force is larger than \SI{30}{\nano\newton}, significantly large areas are cleared of bubbles and wrinkles. This is visible both on the capped and uncapped MoS$_2$ of Fig.\ \ref{fig:Fig1_afm_monolayer}(c). Notably, it is easier to achieve bubble sweeping in the non-covered MoS$_2$ region. Previous studies have reported that the presence of a top hBN layer increases the critical force required to sweep bubbles away with an AFM tip \cite{rosenberger2018nano,chen2021tip,kim2019reliable}. In particular, bubbles larger than \SI{1}{\micro\metre} in size tend to remain in their position, even for forces in the micronewton regime \cite{kim2019reliable}. Reaching these high forces more likely rips apart the bubbles damaging the heterostructure \cite{kim2019reliable}. This can be appreciated in the movie available in the SI for the same sample as in Fig.\ \ref{fig:Fig1_afm_monolayer}, showing the topographic evolution during the ironing process for forces up to \SI{200}{\nano\newton}.} 

{While the cleaning effect is apparent at low forces when comparing Fig.\ \ref{fig:Fig1_afm_monolayer}(a) and (b), and is primarily attributed to the removal of residuals from the top surface, it is not anticipated to significantly affect the optical properties of the heterostructure, as has been shown to be the case for transport characteristics in similar systems \cite{kim2019reliable}. To obtain more insight on the cleaning capabilities in the higher force regime, we conducted a quantitative analysis of the evolution of roughness and bubble morphology in the capped region of Fig.\ \ref{fig:Fig1_afm_monolayer}, for gradually increasing applied forces. Despite the fact that large bubbles were not entirely removed, their shape underwent a noticeable transformation, as depicted in Fig.\ \ref{fig:Fig1_afm_monolayer}(e), where profiles taken prior to and after the AFM treatment are compared for the selected bubble (dotted circle in Fig.\,\ref{fig:Fig1_afm_monolayer}(a)) Moreover, both the height and volume evolution as a function of the applied force can be plotted as illustrated in Fig.\ \ref{fig:Fig1_afm_monolayer}(f) and (g). We note that upon ironing both the height and volume of analyzed bubbles increase (see also Fig.\ \ref{Sfig:AFM_SI_Monolayer}). This can be explained by the fact that at these forces, small bubbles can be displaced and adhere to the larger ones, flattening small corrugations at the hetero-interface in areas between the blisters. This impacts the quality of the interfaces between the layers composing the heterostructures in the areas cleansed of the smallest bubbles, promoting a more intimate contact between the layers. This is expected to enhance the optical quality of the sample, similar to the improved transport characteristics demonstrated in prior work \cite{kim2019reliable}.}
 
{To prove that 
we investigate now a series of three MoS$_2$/MoSe$_2$ heterostructures.} In Fig.\ \ref{fig:Fig2_optical}(a), we present the optical image of sample A (for sample B and sample C see SI). It consists of a MoS$_2$ and MoSe$_2$ monolayers sandwiched between hBN layers (for structural data see table\ \ref{Sfig:table} in the SI). 
The twist angle between monolayers for all structures is determined by second harmonic generation measurements (SI Fig.\ \ref{Sfig:shg}). The results for both tilt angles (\SI{0}{\degree} and \SI{60}{\degree}) are qualitatively the same unless otherwise stated. In the main text, we focus mainly on samples \,A and B (for samples\,C see SI).   

The black dashed triangle in Fig.\ \ref{fig:Fig2_optical}(a) denotes the heterostructure region, where the two monolayers overlap. 
A typical low temperature ($T = \SI{4.5}{\K}$) photoluminescence (PL) spectrum from the heterostructure region is presented in Fig.\ \ref{fig:Fig2_optical}(b). We can distinguish features related to MoSe$_2$ and MoS$_2$ intralayer excitons, at $\sim \SI{1.62}{\eV}$ and $\sim \SI{1.9}{\eV}$ \cite{cadiz2017excitonic}, respectively. These transitions are also visible in the reflectance spectrum of the heterostructure, where we also note an additional feature, corresponding to the MoSe$_2$ B-exciton transition \cite{ross2013electrical}.  On the low-energy side, a strong PL peak, visible only in the heterostructure region, corresponds to the interlayer exciton\cite{baranowski2017probing, mouri2017thermal,  zhang2018moire, surrente2018intervalley} (see also SI Figs.\,\ref{Sfig:sampleB} and \ref{Sfig:sampleC}).

 As shown in Fig.\,\ref{fig:Fig2_optical}(c), {we ironed three square-shaped areas, using an AFM tip in contact mode with 30, 60 and 80\,nN pressing force (for the other samples see SI Tab.\ \ref{Sfig:table}, Fig.\ \ref{Sfig:sampleB}, and Fig.\ \ref{Sfig:sampleC}), resulting in an applied pressure of approximately $4-\SI{6}{\giga\pascal}$ (Sec.\,II of SI). The ironing procedure considerably improves the flatness of the heterostructure as evident in the inset of Fig.\ \ref{fig:Fig2_optical}(c) which shows a close-up view of the corner of the ironed area of the heterostructure. 
 The surface of the ironed part shows the removal of small bubbles and surface adsorbates, and local flattening in between the large bubbles. The presence of bubbles in the non-ironed part of the sample affected the optical response as their in-plane distance was lower than the typical spatial resolution of a $\mu$PL setup.} 

\begin{figure}
    \centering
    \includegraphics[width=1\linewidth]{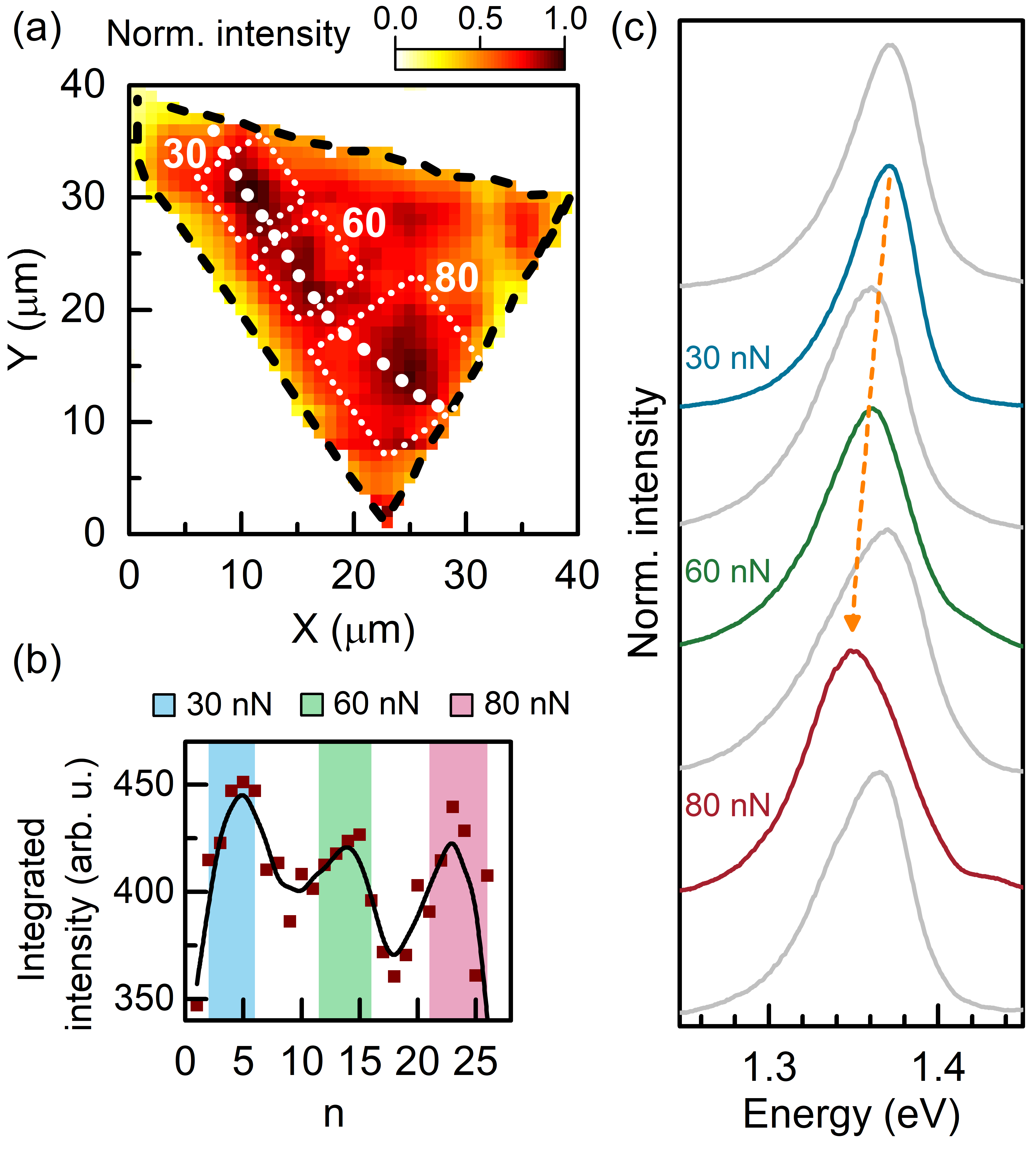}
    \caption{
    (a)  PL intensity map for the interlayer exciton, showing higher intensity hotspots for the ironed regions. (b) The evolution of the PL intensity along the cross-section is indicated by the white dashed line. A clear enhancement of the PL intensity in ironed part (indicated by the shaded area) is visible. (c) Representative PL spectra along the cross-section show the change of the PL line shape in ironed part. Grey spectra are from the non-ironed region along the cross-section direction.   
    }
    \label{fig:Fig3_maps}
\end{figure}

\begin{figure*}
    \centering
    \includegraphics[width=1\linewidth]{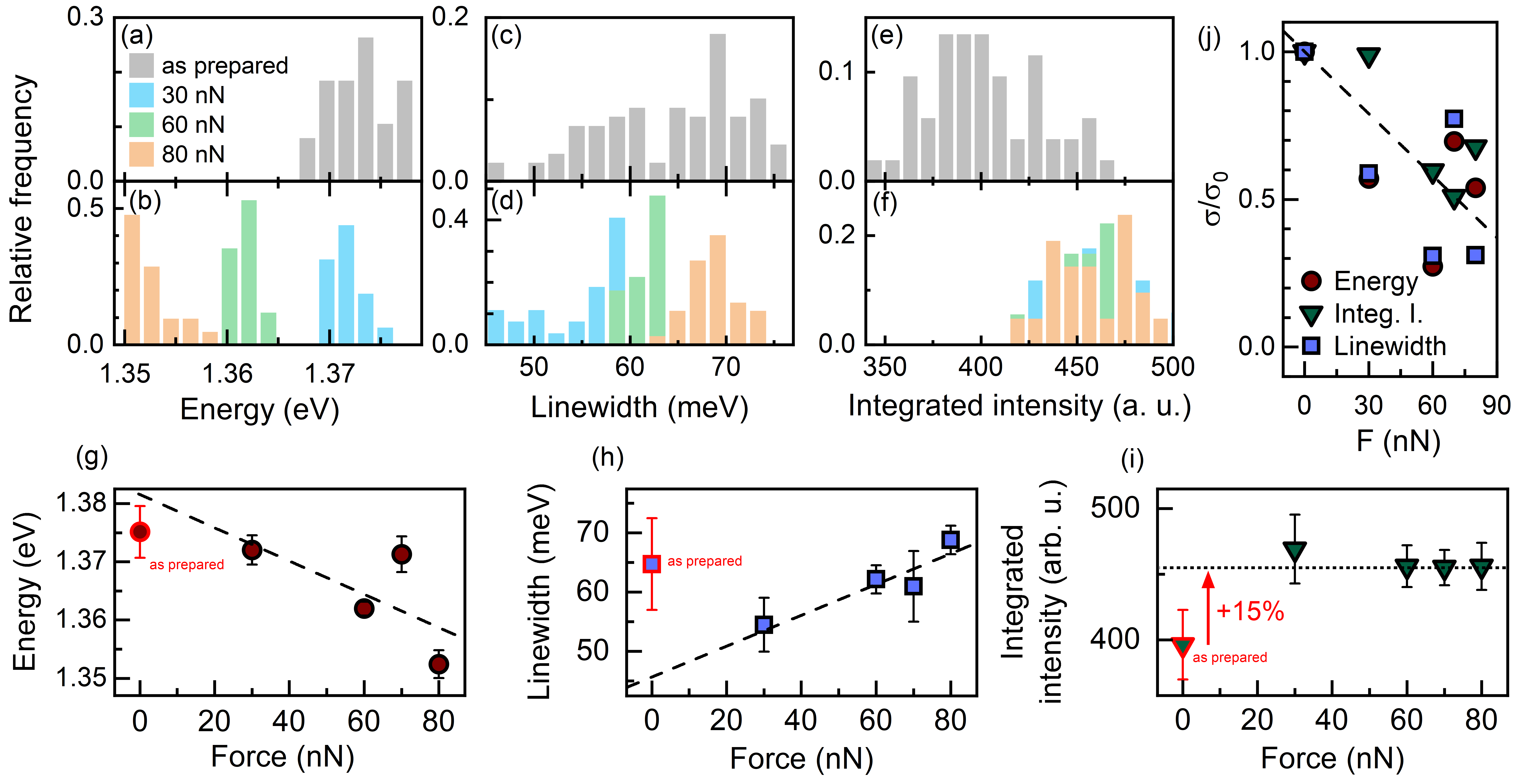}
    \caption{(a-f) Histograms showing the statistical distribution of PL energy, line width and intensity for non-ironed (top panels) and ironed (bottom panels) areas of the heterostructure. (g-i) Average values of interlayer exciton transition energy, line width and PL intensity as a function of ironing force. {Dashed lines are linear fits.} (j) Interlayer exciton transition energy, line width and PL intensity standard deviation $\sigma$ normalised to standard deviation of the corresponding quantity measured on the non-ironed areas $\sigma_0$ as a function of the ironing force. {Dashed line stands for global linear fit of the relative standard deviation of all the parameters}. The fitting procedure in panels (g) and (h) exclude the \SI{0}{\nano\newton} data point{, which corresponds to the as-prepared heterostructure, over which we have no control}.}
    \label{fig:Histograms}
\end{figure*}

Fig.\ \ref{fig:Fig3_maps}(a) presents the spatial dependence of the interlayer exciton PL integrated intensity in the heterostructure area. We observe that, within the ironed regions indicated by the white squares, the PL intensity is enhanced compared with the non-ironed part of the heterostructure. This is especially visible when taking a cross-section through all three ironed regions along the white dotted line, shown in  Fig.\ \ref{fig:Fig3_maps}(b). The integrated PL intensity exhibits local maxima on the ironed parts (shaded areas). In between the ironed regions, the peak intensity drops. Apart from the enhanced intensity, with increasing ironing force the interlayer transition red shifts, broadens and its line shape changes. PL spectra from non-ironed parts exhibit a low energy tail while in the ironed parts line shape is more symmetric with increasing ironing force, and for the highest forces employed, a new feature develops on the high energy side, as illustrated in Fig.\ \ref{fig:Fig3_maps}(c), where representative spectra taken along the cross-section are presented.

To corroborate these observations and to provide a deeper insight into the evolution of interlayer exciton characteristics, we have performed a statistical analysis of the PL spectrum measured at different spots on the heterostructure. The histograms in Fig.\ \ref{fig:Histograms}(a-f) show the distribution of the PL peak energy, line width and intensity for the different forces employed to iron the heterostructure. The ironing results in a red shift of the interlayer exciton transition, as visible in Fig.\ \ref{fig:Histograms}(a,b,g). This red shift is roughly proportional to the ironing force, with a coefficient of $-286 \pm \SI{225}{\micro\eV / \nano\newton}$. The average broadening of the emission line increases with a coefficient of $0.26 \pm \SI{0.07}{\micro\eV / \nano\newton}$, as we discuss more in detail below. Finally, on average the PL intensity in the ironed part increases by about 15\% compared to non-ironed areas. {These results are explained if we consider the more intimate contact between the layers in the heterostructures, due to the reduced density of bubbles we obtain after the AFM treatment. Quantitative estimation of this effect is however complicated by the absence of more detailed structural information, available with transmission microscopy techniques. Moreover, in a MoS$_2$/MoSe$_2$ heterostructure, the interlayer exciton transition is predominantly a strongly hybridized, momentum-indirect exciton \cite{brem2020hybridized,su2016bandgap}, which makes the quantitative analysis of the impact of the interlayer distance very challenging.}

Most importantly, the distributions of all three parameters of interest are narrower after ironing, as evident from the direct comparison between Fig.\ \ref{fig:Histograms}(a,c,e) and Fig.\ \ref{fig:Histograms}(b,d,f), which represent the non-ironed and ironed distributions, respectively. The standard deviation of all the parameters decreases after ironing as shown in Fig.\ \ref{fig:Histograms}(j), where we show the dependence of the standard deviation $\sigma$ of a quantity measured in the ironed area normalised by the standard deviation of the same quantity $\sigma_0$ estimated in non-ironed areas as a function of the ironing force. Our analysis demonstrates that the decreased RMS roughness of the ironed areas {and a reduced density of bubbles and wrinkles} is accompanied by a considerable narrowing (50\% decrease of $\sigma/\sigma_0$) of the statistical distribution of the interlayer exciton energy, line width and intensity along with a general increase of the PL intensity when forces of $60-\SI{80}{\nano\newton}$ are used to iron the heterostructure. This demonstrates that reducing the interface roughness and {locally improving the quality of the interface of the heterostructure} is key to achieving spatially uniform optical spectra.

\begin{figure}
    \centering
    \includegraphics[width=0.9\linewidth]{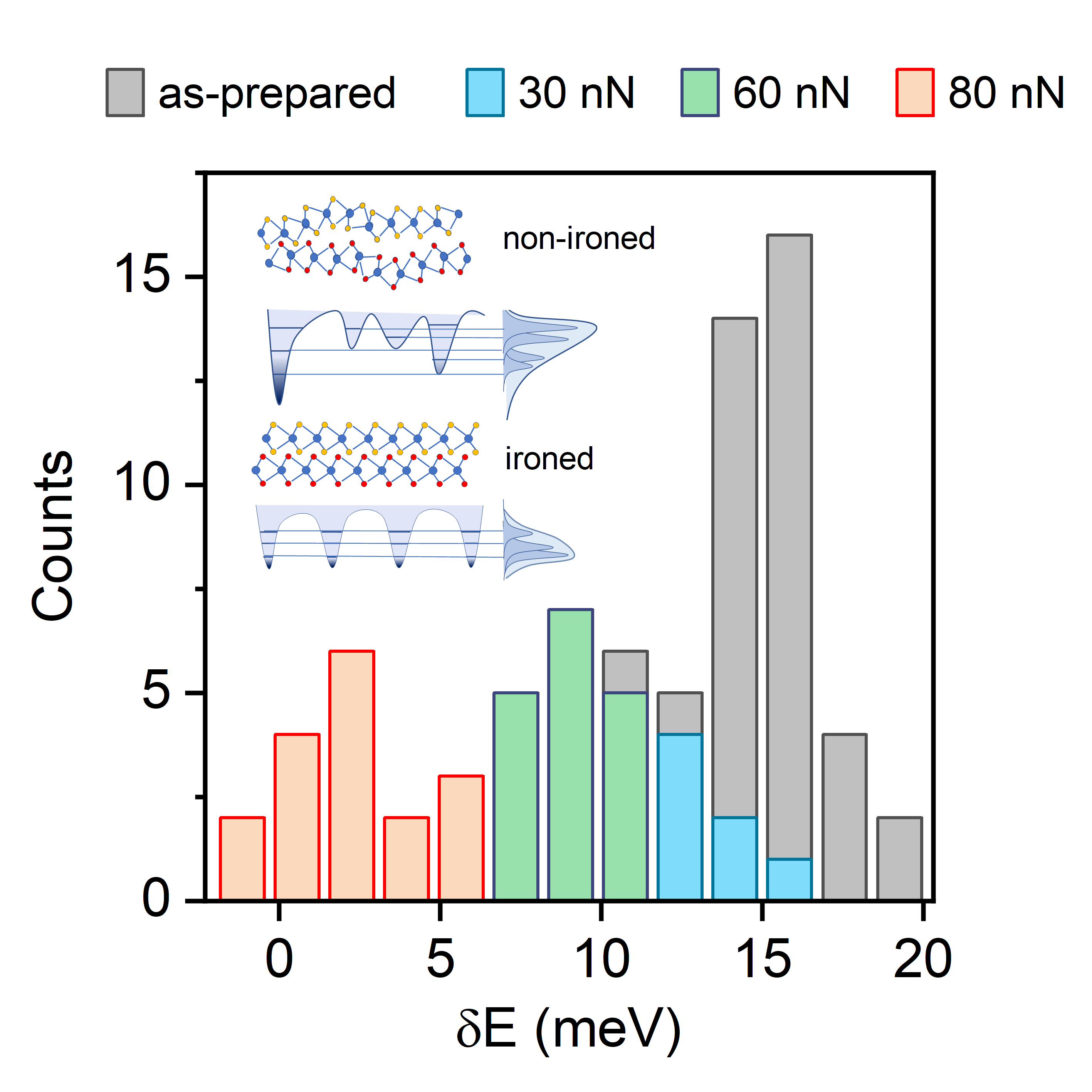}
    \caption {Evolution of the difference $\delta E$ between the maximum of PL peak and PL center of mass as a function of ironing force. A positive $\delta E$ indicates a spectral tail on the low energy side, while negative $\delta E$ indicates a high energy tail. The inset is a schematic illustration of the evolution of the moir{\'e} potential and PL spectra upon ironing. The enhanced proximity of the layers in the flattened structure increases the amplitude of the moir\'e potential, which also becomes more uniform, and on average deeper, therefore it can accommodate more states.}   
    \label{fig:Fig5_force}
\end{figure}


{The observed changes in the interlayer exciton emission properties after ironing are consistent with the enhancement of the moir{\'e} effects \cite{tran2019evidence, wu2018theory, yu2017moire}, due to the improved proximity of the layers and flattening. The red shift of the interlayer exciton transition upon ironing can be understood as an effect of the increased interlayer exciton binding energy and/or band hybridization \cite{brem2020hybridized}, due to a reduction of the interlayer distance. Simultaneously, the improved proximity of the layer together with local flattening deepens the moir\'e trapping potential \cite{tran2019evidence, wu2018theory, yu2017moire} which also enhances redshift of the interlayer exciton.} {The ironing procedure has implications beyond proximity enhancement. Specifically, it is also possible that this procedure serves to mitigate local strain fluctuations and thus increase the moir\'e  effect(see inset of Fig.\ \ref{fig:Fig5_force}).}

The enhancement of the moir\'e effect can also explain the surprising, increased broadening of the interlayer exciton PL spectrum  with the increasing ironing force. However, at the same time, the line width distribution is narrower. This suggests that the increasing broadening is not related to the disorder, but results from the evolution of the intrinsic properties of the interlayer exciton. This is demonstrated in the evolution of the interlayer exciton PL line shape upon ironing as shown in Fig.\ \ref{fig:Fig3_maps}(c). On the non-ironed part, the PL spectrum exhibits a low energy tail, which is a typical indication of disorder in semiconductors \cite{chichibu2001localized, skolnick1986investigation, wright2017band, buyanova1999mechanism, nagler2017interlayer}. In the ironed areas, the PL spectrum is more symmetric or, in some cases, a high-energy shoulder appears (see Fig.\ \ref{fig:Fig3_maps}(c)). To describe the line shape evolution quantitatively, we analyse how the difference $\delta E$ between the maximum of the PL peak ($E_{\textrm{max}}$) and the center-of-mass energy (i.e. $\delta E = E_{\textrm{max}}-\frac{\sum E \cdot I(E)}{\sum I(E)}$) changes as a function of the ironing force. $\delta E$ describes the distribution of the PL intensity with respect to the peak energy. A positive $\delta E$ indicates a spectral tail at the low energy side, whereas negative $\delta E$ is related to the presence of a high energy shoulder. For non-ironed regions the distribution for $\delta E$ is centred around $\sim \SI{15}{\milli\eV}$ (Fig.\,\ref{fig:Fig5_force}). With ironing $\delta E$ decreases and approaches \SI{0}{\milli\eV}, and, for higher forces, it becomes negative, confirming the change of PL lineshape. Again, we expect that this stems from the enhanced coupling between monolayers building the heterostructure after ironing. This is schematically presented in Fig.\ \ref{fig:Fig5_force}. After ironing, the potential landscape becomes more uniform (as attested by reduced variation of full width half maximum) and the increased layers coupling deepens the moir{\'e} potential. This facilitates the bounding of extra states, which contribute to the emission on the high energy side of the PL peak \cite{wu2018theory, tran2019evidence}. This explains the apparent contradiction of an improved spatial uniformity of the line width, observed in conjunction with a constant (or slightly increasing) average value of line width which does not decrease after ironing.

{In conclusion, we have investigated the effect of AFM ironing on the optical quality of TMD heterostructures. We identify two classes of effects. At low forces, the surface roughness of the heterostructure initially increases and then decreases due to a ``sweeping-out'' of surface adsorbates. 
At forces above \SI{30}{\nano\newton}, ironing moves small, more mobile bubbles and merges them into larger bubbles.
We show that AFM ironing can significantly improve the optical quality heterostructures. It improves the homogeneity of the interface (enhancing local flattening, and reducing corrugations and strain fluctuations), leading to more uniform optical spectra and enhanced interlayer coupling and moir{\'e} pattern impact. 
Our findings demonstrate that AFM ironing is a promising method to unveil the intrinsic moir{\'e} physics in TMD heterostructures, which suffer from low reproducibility and inconsistent results related to low interface and surface quality. Therefore, AFM ironing, in addition to hBN encapsulation, could be a crucial post-stacking step to improve the quality of TMD stacks.}  

\section{Methods}
A description of sample preparation, AFM and optical spectroscopy setups can be found in SI.

\section{Acknowledgements}

We would like to thank Thomas Pucher and Alvaro Rodriguez Rodriguez whose timely help with the exfoliation of TMD flakes helped us a lot in the final stages of the work. This work received funding from the European Union’s Horizon 2020 research and innovation program under grant agreements 956813 (2Exciting) and 755655 (ERC-St G 2017 project 2D-TOPSENSE). M.B. acknowledges National Science Centre Poland within the SONATA BIS program (Grant No.\ 2020/38/E/ST3/00194). Funding was also received from the Ministry of Science and Innovation (Spain) through the project PID2020-115566RB-I00 and the EU FLAG-ERA project ``To2Dox'' under the program PCI2019-111893-2. This study has been partially supported through the EUR grant NanoX n$^\circ$ ANR-17-EURE-0009 in the framework of the ``Programme des Investissements d’Avenir''. M.D.\ acknowledges the support from the Polish National Agency for Academic Exchange (grant no.\ BPN/BKK/2021/1/00002/U/00001). K.W.\ and T.T.\ acknowledge support from JSPS KAKENH I (Grant Numbers 19H05790, 20H00354 and 21H05233)

\bibliography{Biblio}

\end{document}